# Probing the Spin Pumping Mechanism: Exchange Coupling with Exponential Decay in $Y_3Fe_5O_{12}$/barrier/Pt Heterostructures

C. H. Du[1,†], H. L. Wang[1,†], Y. Pu[1], T. L. Meyer[2], P. M. Woodward[2], F. Y. Yang[1,*], and P. C. Hammel[1,*]

[1]Department of Physics, The Ohio State University, Columbus, OH, 43210, USA

[2]Department of Chemistry and Biochemistry, The Ohio State University, Columbus, OH, 43210, USA

[†]These authors made equal contributions to this work

[*]Emails: fyyang@physics.osu.edu; hammel@physics.osu.edu

**Ferromagnetic resonance (FMR) driven spin pumping of pure spin currents from a ferromagnet (FM) into a nonmagnetic material (NM) promises new spin-functional devices with low energy consumption [1-6]. The mechanism of spin pumping is under intense investigation and it is widely believed that exchange interaction between the FM and NM is responsible for this phenomenon [2-5, 7]. We observe a thousand-fold exponential decay of the spin pumping from 20-nm thick $Y_3Fe_5O_{12}$ (YIG) films to platinum across insulating barriers, from which the exponential decay lengths of 0.16 and 0.23 nm are extracted for oxide barriers with band gaps of 4.93 eV and 2.36 eV, respectively. This prototypical signature of quantum tunneling through a barrier underscores the importance of exchange coupling for spin pumping and reveals its dependence on the characteristics of the barrier material.**

Generation and manipulation of spin currents is key to spintronic applications [1]. FMR driven spin pumping has been demonstrated to inject a pure spin current through angular momentum transfer from an FM to an adjacent NM [2-6]. General belief assumes a dynamic coupling via exchange interaction between the precessing magnetization (**M**) of the FM and the conduction electrons of the NM at the NM/FM interface. This mechanism will lead to a short, atomic-scale coupling that decays exponentially with separation between the FM and NM. However, this has not been experimentally confirmed, partially due to the large dynamic range needed to measure such rapidly decaying spin pumping signal [6, 8]. Our recent demonstration of large spin pumping in Pt/YIG bilayers with mV-level inverse spin Hall effect (ISHE) voltage, $V_{ISHE}$, offers a material platform with signal-to-noise ratio sufficient to quantitatively characterize the coupling range, enabling detailed insights into the spin pumping mechanism. We use three different barrier materials, including two oxide insulators and Si, to systematically investigate the barrier thickness $t$ dependence of spin pumping in Pt/barrier($t$)/YIG heterostructures. The clear exponential decays of ISHE voltage with characteristic lengths of ~0.2 nm for the oxide barriers provide decisive evidence for the predicted exchange coupling model for spin pumping.

Our experiments utilize YIG thin films grown on (111)-oriented $Gd_3Ga_5O_{12}$ (GGG) substrates by an off-axis sputtering technique we developed for epitaxial film growth of complex materials [9-12]. Figure 1a shows a high resolution x-ray diffraction (XRD) scan of a 20-nm YIG film near the YIG (444) peak with clear Laue oscillations, indicating high uniformity throughout the film. The XRD rocking curve in the inset to Fig. 1a gives a full width at half maximum (FWHM) of 0.0185°, demonstrating excellent crystalline quality. Figure 1b shows a representative FMR derivative spectrum for a 20 nm YIG film taken at radio-frequency (rf) $f =$ 9.65 GHz and microwave power $P_{rf} = 0.2$ mW with an in-plane magnetic field, from which the peak-to-peak linewidth ($\Delta H$) of 10 Oe is obtained [13].

Spin pumping measurements are conducted at room temperature using a 5-nm thick Pt layer on 20-nm YIG films, as illustrated in Fig. 1c. The samples with a width of ~1 mm and a length of ~5 mm are placed in the center of an FMR cavity and a DC magnetic field (**H**) is applied in the *xz*-plane. At resonance, the precessing magnetization transfers angular momentum from YIG to the conduction electrons in Pt by means of dynamical coupling [6, 14], generating a pure spin current, **J**$_s$, in Pt along the *z*-axis with polarization (**σ**) parallel to the YIG magnetization. In the Pt layer, **J**$_s$ is converted into a net charge current, **J**$_c$∝**J**$_s$×**σ**, via inverse spin Hall effect [15-18], resulting in an ISHE voltage along the *y*-axis. Figure 1d shows the $V_{ISHE}$ vs. $H$ spectra of a Pt/YIG bilayer at $\theta_H = 90°$ and 270° (field in plane) at $P_{rf} = 200$ mW. The peak value of $V_{ISHE} = 1.0$ mV is generated at the FMR resonant condition given by the FMR derivative spectrum in Fig. 1e. As **H** is reversed from $\theta_H = 90°$ to 270°, $V_{ISHE}$ changes sign and maintains the same magnitude as expected since **σ** changes sign with reversal of **M** (**M** ∥ **H** since $H$ exceeds $4\pi M_s$ at resonance), resulting in a reversed sign of the $V_{ISHE}$.

Figure 1f shows the angular dependence of normalized $V_{ISHE}(\theta_H)$ for the Pt/YIG bilayer, which agrees with the green curve of a simple $\sin\theta_H$, confirming that the observed voltage signal comes from FMR spin pumping due to,

$$V_{ISHE} \propto \mathbf{J_s}\times\mathbf{\sigma} \propto \mathbf{J_s}\times\mathbf{M} \propto \mathbf{J_s}\times\mathbf{H} \propto \sin\theta_H. \tag{1}$$

The ISHE voltage of the Pt/YIG bilayer at $\theta_H = 90°$ in Fig. 1g is proportional to power $P_{rf}$ for 0.2 mW $< P_{rf} <$ 200 mW, indicating that the observed mV-level $V_{ISHE}$ is still in the linear regime.

To characterize the coupling range of spin pumping, we insert three different thin, insulating barriers, Sr$_2$GaTaO$_6$ (SGTO) with a band gap $E_g = 4.93$ eV, Sr$_2$CrNbO$_6$ (SCNO) with $E_g = 2.36$ eV and amorphous Si between Pt and YIG as illustrated in Fig. 1c (see Supplementary Information for detailed characteristics of double perovskite Sr$_2$GaTaO$_6$ and Sr$_2$CrNbO$_6$).

Figures 2a-2f show the spin pumping and FMR derivative absorption spectra of Pt/barrier/YIG structures with 0.5-nm $Sr_2GaTaO_6$, $Sr_2CrNbO_6$ and Si barriers, which reduce the ISHE voltage to 20, 100 and 440 $\mu$V, respectively. This sensitivity of the decay rate of $V_{ISHE}$ will be discussed later.

Figures 2g-2i show the angular dependence of normalized $V_{ISHE}$ for Pt/barrier(0.5nm)YIG with the three barriers. The sinusoidal angular dependence is characteristic of ISHE [see Eq. (1)], confirming that the observed signals are due to spin pumping, not artifacts due to thermoelectric or magnetoelectric effects such as anisotropic magnetoresistance (AMR) [19, 20]. Figures 2j-2l show the rf power dependence of $V_{ISHE}$ from $P_{rf}$ = 0.2 mW to 200 mW with an in-plane field ($\theta_H$ = 90°) for the three samples, all showing linear relationship between $P_{rf}$ and $V_{ISHE}$ [21].

To understand the systematic behavior of spin pumping across a thin insulating barrier, we plot the dependence of $V_{ISHE}$ (normalized by the Pt/YIG samples with direct contact) on the barrier thickness in Figs. 3a-3c for Pt/$Sr_2GaTaO_6$(t)/YIG, Pt/$Sr_2CrNbO_6$(t)/YIG and Pt/Si(t)/YIG heterostructures. Four representative $V_{ISHE}$ vs. $H$ spectra for various barrier thicknesses are shown in Figs. 3e-3f for each series. The mV scale of the Pt/YIG ISHE voltage allows us to observe dramatic, thousand-fold changes in $V_{ISHE}$. As the barrier thickness increases, $V_{ISHE}$ exhibits a clear exponential decay for all three barrier materials and eventually falls below the noise level at $t$ = 2 nm for $Sr_2GaTaO_6$ and $Sr_2CrNbO_6$ barriers and at $t$ = 5 nm for Si barrier. From a least-squares linear fit shown in the semi-log plots in Figs. 3a-3c, we obtain an exponential decay length, $\lambda$ = 0.16, 0.23 and 0.74 nm for $Sr_2GaTaO_6$, $Sr_2CrNbO_6$ and Si barriers, respectively, following,

$$V_{ISHE} = V_{ISHE}(t=0)e^{-t/\lambda}. \tag{2}$$

Given $V_{ISHE} \propto J_s$, the exponential decay of $V_{ISHE}$ indicates that the pure spin current generated in Pt also decreases exponentially with *t*, a signature behavior of exchange coupling between FM and NM separated by an insulating barrier. This result is the first direct quantitative evidence of the exchange coupling model for spin pumping.

The exponential dependence of spin pumping on barrier thicknesses can be explained by a process in which the wavefunction of the conduction electrons in Pt tunnels through the barrier, couples with the precessing magnetization of YIG through exchange interaction, and acquires spin polarization via spin-dependent scattering at the barrier/YIG interface [2-5]. At a NM/FM interface, a spin current can be generated either by transmission of spin-polarized electrons from the FM into NM, or by spin-dependent scattering of the conduction electrons in NM at the interface. Given that YIG is an insulator, it is unlikely that spin-polarized electrons flow from YIG into Pt. Consequently, the dominant mechanism is the spin-dependent scattering of conduction electrons in Pt at the Pt/YIG interface via exchange interaction with the precessing magnetization of YIG. When the Pt and YIG are separated by a thin barrier within a tunneling distance, the wavefunction of the conduction electrons in Pt can tunnel through the barrier and couple with YIG, and the tunneling probability depends sensitively on the barrier height.

At the interface between a metal and an insulator or a semiconductor, the relevant barrier is typically the Schottky barrier, $\Phi_B$, which depends on the work function of the metal and the electron affinity, charge carrier type and concentration of the insulator/semiconductor. Schottky barrier heights for $Pt/Sr_2GaTaO_6$ and $Pt/Sr_2CrNbO_6$ are not known; here we estimate the values of $\Phi_B$ for $Pt/Sr_2GaTaO_6$ and $Pt/Sr_2CrNbO_6$ based on published results on metal/perovskite Schottky junctions and correlate them to the extracted decay lengths. It was reported that in $Au/SrTiO_3$ (Nd-doped, carrier density $10^{17} - 10^{18}$ cm$^{-3}$) Schottky junctions [22], the Schottky barrier height is in the range of 1.4 – 1.7 eV, about half of the SrTiO$_3$ band gap. At lower carrier concentration, $\Phi_B$ is expected to remain in this range [22]. Since Au and Pt have similar work functions (5.47 eV for Au and 5.64 eV for Pt) and both Sr$_2$GaTaO$_6$ and Sr$_2$CrNbO$_6$ are Sr-based perovskites [23], it is reasonable to expect $\Phi_B$ in Pt/Sr$_2$GaTaO$_6$ and Pt/Sr$_2$CrNbO$_6$ to be half of their barrier band gaps as well. Using band gaps of 4.93 eV (Sr$_2$GaTaO$_6$) and 2.36 eV (Sr$_2$CrNbO$_6$) determined by optical absorption (see Supplementary Information for details), we estimate $\Phi_B$ = 2.5 and 1.2 eV for Pt/Sr$_2$GaTaO$_6$ and Pt/Sr$_2$CrNbO$_6$, respectively.

For a finite rectangular potential barrier, as illustrated in Figs. 4a and 4b for Sr$_2$GaTaO$_6$ and Sr$_2$CrNbO$_6$ barriers, respectively, the tunneling transmission coefficient $D$ of electrons is determined by the barrier height $\Phi_B$ and width $t$ [24]:

$$D \propto \exp\left[-\frac{2t}{\hbar}\sqrt{2m\,\Phi_B}\right], \qquad (3)$$

where $m$ is the effective mass and $\hbar$ is Planck's constant. Since $V_{\text{ISHE}} \propto J_s \propto D$, Eqs. (2) and (3) imply $1/\lambda \propto \sqrt{\Phi_B}$, as shown in Fig. 4c for Pt/Sr$_2$GaTaO$_6$($t$)/YIG and Pt/Sr$_2$CrNbO$_6$($t$)/YIG. The two data points align with the origin, providing further evidence for the exchange coupling model in spin pumping and the role of barrier materials in quantum tunneling.

For Si barriers, the decay length of 0.74 nm is much larger than the 0.16 and 0.23 nm for samples with oxide barriers. Si has a smaller band gap (1.1 eV) than the two oxide barriers, thus we expect a larger decay length. If we use $\Phi_B$ = 0.55 eV for Pt on amorphous Si, we estimate a decay length of 0.34 nm using the same rectangular potential barrier model, smaller than observed, indicating a barrier height smaller than half of the Si band gap as we assumed for the oxide barriers. This is not unexpected, however, since the Schottky barrier heights of metal/Si junctions are sensitive to the doping type and carrier concentration. In addition to the mechanism already discussed, we should also consider the possibility that there may be carriers in the Si barrier allowing either an indirect exchange process or spin diffusion through the barrier [25]. Further investigation of the characteristics of the barriers in spin pumping is needed to obtain better insights into the spin pumping mechanisms.

Experimental observation of a clear exponential decay of dynamic spin pumping provides decisive evidence for, and quantitative understanding of a fundamental spin pumping mechanism. This result points to the important ability to tune characteristics of spin functional devices and reveal new phenomena.

**Acknowledgements**

This work is supported by the Center for Emergent Materials at the Ohio State University, a NSF Materials Research Science and Engineering Center (DMR-0820414) (HLW, YP, and FYY) and by the Department of Energy through grant DE-FG02-03ER46054 (PCH). Partial support is provided by Lake Shore Cryogenics Inc. (CHD) and the NanoSystems Laboratory at the Ohio State University.

**Figure Captions:**

**Figure 1. a** Semi-log $\theta$-$2\theta$ XRD scan of a 20-nm thick YIG film on GGG(111), which exhibits clear Laue oscillations corresponding to the film thickness. Inset: rocking curve of the YIG (444) peak. **b** Representative room-temperature FMR derivative spectrum of a 20-nm YIG film with ***H*** in plane at $P_{rf}$ = 0.2 mW, which gives a peak-to-peak linewidth of 10 Oe. **c** Schematics of experimental setup for ISHE voltage measurements for Pt/YIG and Pt/barrier/YIG samples. **d** $V_{ISHE}$ vs. *H* spectra and **e** FMR derivative spectrum of a Pt(5nm)/YIG(20nm) bilayer at $P_{rf}$ = 200 mW. **f** Angular dependence of normalized $V_{ISHE}$ of the Pt/YIG bilayer (the green curve is a simple calculation of $\sin\theta_H$). **g** rf power dependence of $V_{ISHE}$ with a least-squares fit.

**Figure 2.** $V_{ISHE}$ vs. *H* spectra of Pt/barrier/YIG multilayers with 0.5-nm thick **a** $Sr_2GaTaO_6$, **b** $Sr_2CrNbO_6$, and **c** Si barriers at $P_{rf}$ = 200 mW, with the corresponding FMR derivative spectra shown in **d, e, f**, respectively. Angular dependencies of the normalized $V_{ISHE}$ in **g, h, i** follow $\sin\theta_H$ (green curves), and the rf power dependencies of $V_{ISHE}$ in **j, k, l** show linear behavior for all three samples.

**Figure 3.** Semi-log plots of $V_{ISHE}$ normalized to the values for Pt/YIG bilayers with direct contact as a function of the barrier thickness of **a** Pt/$Sr_2GaTaO_6$/YIG, **b** Pt/$Sr_2CrNbO_6$/YIG, and **c** Pt/Si/YIG, with four representative $V_{ISHE}$ vs. *H* spectra of various barrier thicknesses for each series shown in **d, e**, and **f**, respectively. The solid lines in **a, b**, and **c** are least-squares fits, which indicate exponential decay of ISHE voltages with decay length $\lambda$ = 0.16, 0.23, and 0.74 nm for Pt/$Sr_2GaTaO_6$/YIG, Pt/$Sr_2CrNbO_6$/YIG, and Pt/Si/YIG, respectively.

**Figure 4.** Schematics of band structures of **a** Pt/$Sr_2GaTaO_6$/YIG and **b** Pt/$Sr_2CrNbO_6$/YIG heterostructures with an estimated Schottky barrier height $\Phi_B$ at half of the bandgap for each material. The blue curves in **a** and **b** illustrate the quantum tunneling of the electron

wavefunction from Pt into the barrier. **c** Inverse of decay length, $1/\lambda$, as a function of $\sqrt{\Phi_B}$ for Pt/Sr$_2$GaTaO$_6$/YIG and Pt/Sr$_2$CrNbO$_6$/YIG. The solid markers are experimental data and the solid line connecting the two points and origin is guide to the eye.

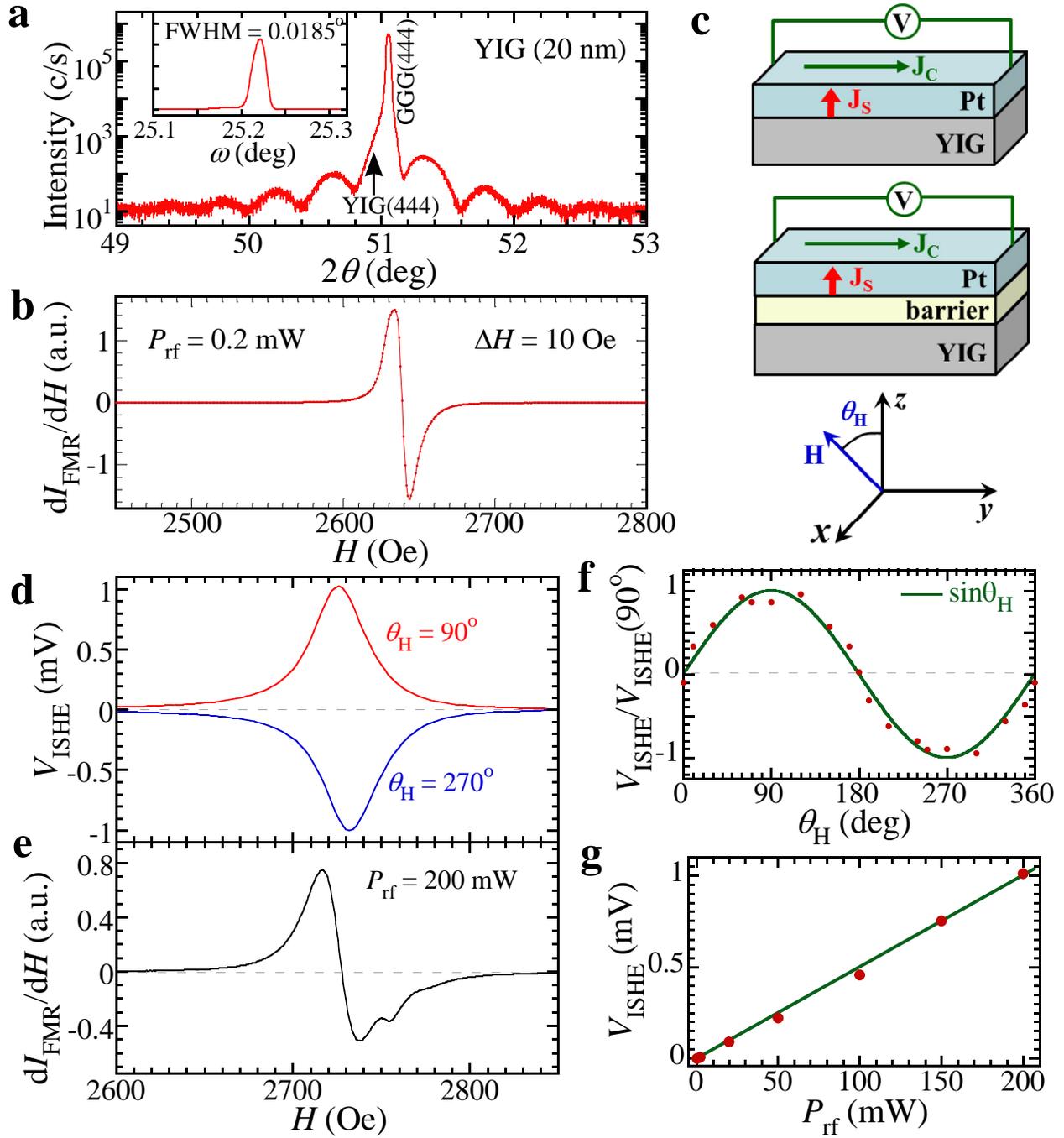

**Fig. 1.**

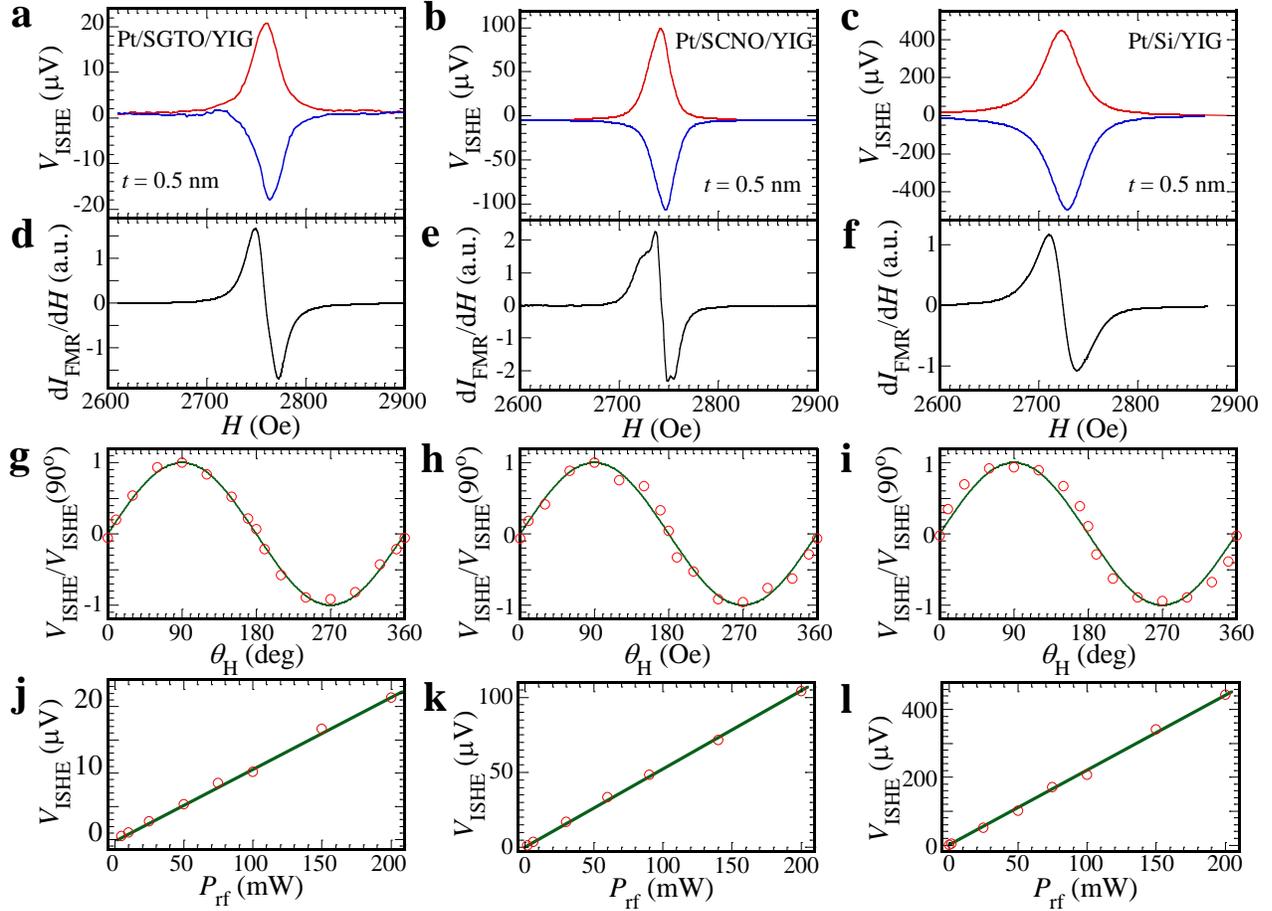

**Fig. 2.**

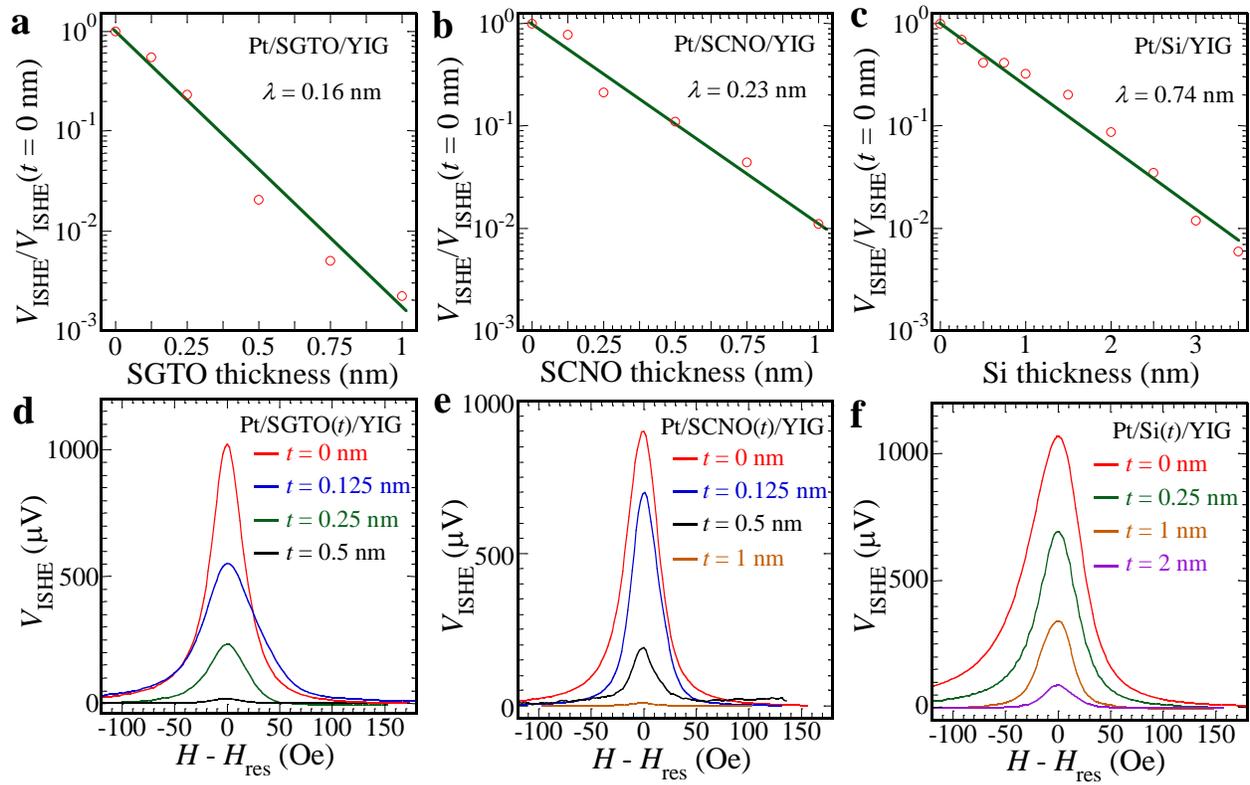

**Fig. 3.**

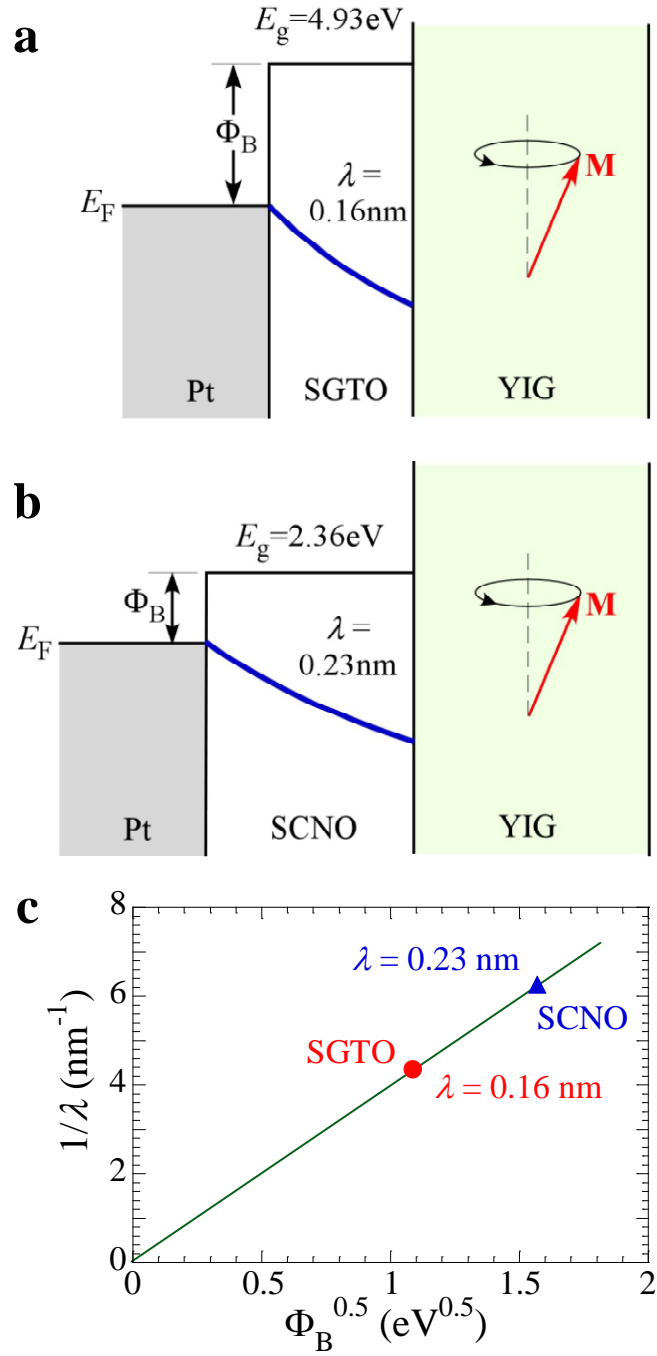

**Fig. 4.**

## Supplementary Information

1. **Growth of $Y_3Fe_5O_{12}$, Pt and Si layers**

Single-crystalline $Y_3Fe_5O_{12}$ (YIG) epitaxial thin films were grown on (111)-oriented $Gd_3Ga_5O_{12}$ (GGG) substrates in an off-axis ultrahigh vacuum (UHV) sputtering system [S1-S3] with a base pressure below $5 \times 10^{-9}$ Torr. The optimal growth conditions include: a total $Ar/O_2$ pressure of 11.5 mTorr with an $O_2$ concentration of 0.15%, a substrate temperature 750 °C, and a radio-frequency (rf) sputtering power of 50 W. The deposition rate of YIG films is 0.33 nm/min. Crystal structure of the films was characterized by a Bruker D8 high resolution x-ray diffraction (XRD) system. Polycrystalline Pt and amorphous Si layers were deposited in the same sputtering system in pure Ar at room temperature using off-axis geometry. The deposition rates of Pt and Si are between 1 and 1.6 nm/min. Both in-situ and ex-situ growths of the Pt layers were carried out to test whether exposing the YIG films to air would affect the Pt/YIG interface quality and the efficiency of spin pumping. However, no difference in the magnitude of the spin pumping signals is observed between the two growth modes.

2. **$Sr_2GaTaO_6$ and $Sr_2CrNbO_6$ barriers**

$Sr_2GaTaO_6$ (SGTO) and $Sr_2CrNbO_6$ (SCNO) are both insulators with the $A_2BB'O_6$ double perovskite structure, which can be viewed as combination of two different single perovskites (similar to $SiTiO_3$). Double perovskites exhibit a broad range of interesting properties due to their complexity and tenability, and have been used in various applications. For example, LSAT is a commonly used single crystal substrate for epitaxial film growth with the composition $(LaAlO_3)_{0.3}(Sr_2AlTaO_6)_{0.7}$ [S4], of which the majority phase ($Sr_2AlTaO_6$) is a double perovskite very similar to the $Sr_2GaTaO_6$ used in this report. $Sr_2CrNbO_6$ has an almost identical lattice constant as $Sr_2GaTaO_6$ ($a$ = 0.788 nm) and have been used as an epitaxial buffer layer for subsequent growth of other double perovskite epitaxial films [S5].

Thin $Sr_2GaTaO_6$ and $Sr_2CrNbO_6$ layers of various thicknesses were deposited on YIG films in the same sputtering system at room temperature by rf sputtering using a power of 50 W. The $Ar/O_2$ sputtering pressure is 12.5 mTorr and the $O_2$ concentration is between 0.15% and 0.20%. The deposition rates are 2.2 and 0.50 nm/min for $Sr_2CrNbO_6$ and $Sr_2GaTaO_6$, respectively.

### 3. Characterization of band gap by UV-visible absorption

Diffuse reflectance (R) spectra of the $Sr_2CrNbO_6$ and $Sr_2GaTaO_6$ samples were collected on an Ocean Optics USB4000-UV-VIS miniature fiber optic spectrometer using a Spectralon® standard. The percent reflectance was transformed using the Kubelka-Munk function [S6, S7], $F(R) = \frac{(1-R)^2}{2R}$, and is shown in Fig. S1. The intersection of the linear fit and photon energy axis gives band gaps of 4.93 eV ($Sr_2GaTaO_6$) and 2.36 eV ($Sr_2CrNbO_6$).

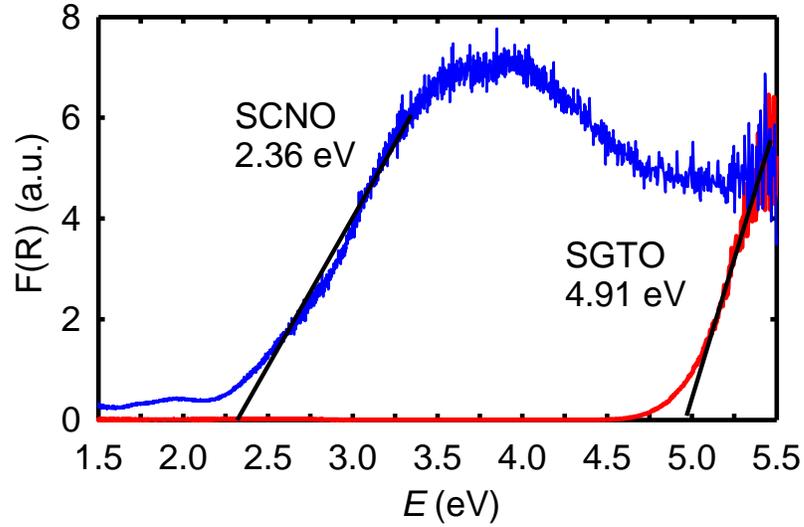

**Figure S1:** UV-visible absorption spectra for $Sr_2GaTaO_6$ (red) and $Sr_2CrNbO_6$ (blue) using the Kubelka-Munk function, $F(R) = \frac{(1-R)^2}{2R}$, where R is the reflectance. From the intersection of the absorption edges and the x-axis, the band gaps of 4.93 and 2.36 eV are obtained for $Sr_2GaTaO_6$ and $Sr_2CrNbO_6$, respectively.